\documentclass[letterpaper, 10pt, conference, compsocconf]{IEEEtran}
\usepackage{color,graphicx,threeparttable,amsmath,textcomp}
\usepackage[official]{eurosym}
\begin{document}

\title{Cellular IoT Traffic Characterization and Evolution}

\author{\IEEEauthorblockN{Benjamin Finley\IEEEauthorrefmark{1}, Alexandr Vesselkov\IEEEauthorrefmark{2}}
\IEEEauthorblockA{Department of Communications and Networking\\
Aalto University, Helsinki, Finland\\
Email: \IEEEauthorrefmark{1}benjamin.finley@aalto.fi,}
\IEEEauthorrefmark{2}alexandr.vesselkov@aalto.fi}

\IEEEoverridecommandlockouts
\IEEEpubid{\makebox[\columnwidth]{978-1-5386-4980-0/19/\$31.00 \copyright~2019~IEEE}
\hspace{\columnsep}\makebox[\columnwidth]{ }}

\maketitle

\begin{abstract}
The adoption of Internet of Things (IoT) technologies is increasing and thus IoT is seemingly shifting from hype to reality. However, the actual use of IoT over significant timescales has not been empirically analyzed. In other words the reality remains unexplored. Furthermore, despite the variety of IoT verticals, the use of IoT across vertical industries has not been compared. This paper uses a two-year IoT dataset from a major Finnish mobile network operator to investigate different aspects of cellular IoT traffic including temporal evolution and the use of IoT devices across industries. We present a variety of novel findings. For example, our results show that IoT traffic volume per device increased three-fold over the last two years. Additionally, we illustrate diversity in IoT usage among different industries with orders of magnitude differences in traffic volume and device mobility. Though we also note that the daily traffic patterns of all devices can be clustered into only three patterns, differing mainly in the presence and timing of a peak hour. Finally, we illustrate that the share of LTE-enabled IoT devices has remained low at around 2\% and 30\% of IoT devices are still 2G only.
\end{abstract}
\begin{IEEEkeywords} 
IoT, M2M, Empirical Measurements, Cellular Network
\end{IEEEkeywords}

\IEEEpeerreviewmaketitle

\section{Introduction}\label{sec:introduction}
According to a recent survey \cite{AnalysysMason2017}, 29\% of companies globally utilize Internet of Things (IoT) devices, suggesting that IoT is moving from hype to reality. Furthermore, IoT is gaining momentum across different industries that use IoT devices in unique ways for solving diverse problems. Despite this variety of IoT verticals and applications, the actual usage of IoT across industries has never been empirically explored. Additionally, the few studies \cite{Marjamaa2012, shafiq2013, Romirer-Maierhofer2015} that have analyzed IoT device usage from commercial cellular networks are relatively old and have only analyzed short timescales (typically less than a few weeks). Thus the evolution of IoT usage over longer timescales remains uninvestigated.

To address these gaps, in this work we analyze a two-year IoT dataset from a major Finnish mobile network operator (MNO) that includes data traffic volumes, customer industry class, and device features. More specifically the analysis focuses on traffic and mobility patterns of IoT devices on the industry-level and on several different timescales. The analysis also covers the evolution of the features and age of the IoT device base over the two years. Overall, the work gives a holistic view of the evolution and current state of IoT usage in a major MNO, thus illustrating the reality instead of the hype. We note that Finland is an early IoT adopter with the 6th most M2M modules per capita of OECD countries (23 per 100 inhabitants) \cite{oecdm2m2017}.

The results of this study are relevant to both researchers and practitioners. In particular, researchers studying the impact of IoT on future cellular networks can use the identified IoT device traffic patterns for improved modeling. Furthermore, providers of IoT connectivity and other services can get a better understanding of the requirements and challenges of IoT devices in different verticals, which will allow them to address customer needs. Finally, knowing typical IoT traffic patterns, companies planning to deploy IoT projects can better evaluate the operating costs that will arise on the network connectivity side.

\section{Related Work}\label{sec:background}
Shafiq et al. \cite{shafiq2013} were the first to analyze IoT\footnote{They denoted such traffic as machine-to-machine (M2M).} data from a commercial cellular network. They examined the traffic generated by more than a million IoT devices over one week in August 2010 and found that such devices are less mobile than smartphones, generate more uplink than downlink traffic, and often have synchronized activity. Ref. \cite{Romirer-Maierhofer2015} confirmed these last two observations by analyzing IoT device data collected over several weeks in 2013. Both studies concluded that the traffic generated by IoT devices significantly differs from smartphones, indicating the need for MNOs to reassess network planning traditionally optimized for smartphone users.

In a more recent study, Andrade et al. \cite{Andrade2017} analyzed the traffic and mobility patterns of one million connected cars on a cellular network in the US. The authors concluded that the data traffic that cars generate differ both from smartphones and other IoT devices, and warned about the potential adverse impact that massive over-the-air firmware updates may have on network performance.

Finally, several studies \cite{Baer2016, Baer2015, Laner2014} similarly analyzed IoT data from a cellular network but with different objectives. They proposed methods for online and offline classification of IoT traffic that would give MNOs a more efficient way of identifying IoT devices compared to the traditional TAC-based (Type Allocation Code) approach. 

\section{Dataset}\label{sec:dataset}
The main dataset of the analysis is a collection of data detail records (DDRs) of devices that use business-focused IoT-specific subscriptions from a major Finnish MNO. The dataset covers a period of 2 years from September 2016 to August 2018. Each record covers a single hour and contains the following fields: anonymized IMSI\footnote{We only refer to devices in this work and we assume a one-to-one relationship between IMSI and device since SIM cards are rarely swapped to different devices. Empirically we find that only 1.6\% of IMSIs were used with multiple devices over the entire period.}, anonymized cell ID, anonymized customer ID (hereafter company ID), device TAC, uplink traffic volume, and downlink traffic volume. If the device had traffic in more than one cell in a given hour then additional records for that hour for each cell were included.

Additionally, the DDR dataset was joined with two other MNO provided datasets: a dataset of device features (from the GSMA device database) for all TACs found in the DDR dataset and a dataset of company industries for each company ID in the DDR dataset. The device feature dataset includes fields such as device model name, release year, and network capabilities (i.e. EDGE, HSPA, LTE, etc), while the company industry dataset is based on the standard Finnish TOL2008\footnote{TOL2008 is based on the EU's classification of economic activities, NACE Rev.2, prescribed in the EC Regulation (EC) no1893/2006} industry classification. For industry-level analyses, we only include industries with a sufficient number of companies to allow for meaningful generalizations. For reference, we list these industries, their acronyms (used in figures), and brief descriptions in Table \ref{tab:industry_descriptions}.

To ensure that only IoT devices were included in the analysis, we first manually checked all unique device models from the dataset and categorized them as IoT, maybe-IoT (typically PCI Express data cards that can be also used in laptops), and non-IoT (typically smartphones and feature phones) based on online research. We then filtered out all non-IoT devices and any device with an invalid TAC code (since in those cases we did not have any device information). This filtering removed 5.7\% of devices.

To give an idea of the full scale of the analysis, the DDR dataset covers hundreds of companies, hundreds of thousands of devices, and tens of millions of records. We also note that for business confidentiality and privacy reasons we normalize some of the numerical results, however this normalization does not change the interpretations or conclusions. Finally, for illustration purposes we use moving averages\footnote{The moving average is essentially a low-pass filter in signal processing.} in several figures to help emphasize longer-term trends and smooth out short-term fluctuations.

\begin{table*}[ht]
\caption{Description of industries based on NACE Rev. 2 \cite{Eurostat2008}}
\label{tab:industry_descriptions}
\begin{tabular}{p{3.8cm}cp{12.2cm}}
\hline
\textbf{Industry (abbreviated)} & \textbf{Acronym} & \textbf{Description} \\ \hline
Administrative and support & AS & Activities supporting general business operations, except professional activities; e.g., rental and leasing, recruitment, security and investigation \\\hline
Electricity and gas & EG & Providing electric power, natural gas, steam, hot water and the like through a permanent infrastructure \\\hline
Information and communication & IC & Publishing activities, including SW; broadcasting; telecommunications and IT activities \\\hline
Manufacturing & MF & Physical or chemical transformation of materials or components into new products, e.g. food, textiles, computers and electronics \\\hline
Professional activities & PA & Activities making specialized knowledge available to users, e.g., consultancy and engineering \\\hline
Transportation & TR & Provision of passenger or freight transport, and associated activities such as terminal and parking facilities, cargo handling, and storage \\\hline
Wholesale and retail trade & WR & Wholesale and retail sale of any goods, including associated operations, such as assembling and packing; repair of motor vehicles and motorcycles \\\hline
\end{tabular}%

\end{table*}
\section{Results}\label{sec:results}

\subsection{Traffic statistics}
\begin{figure}
\centering
\includegraphics[width=\columnwidth]{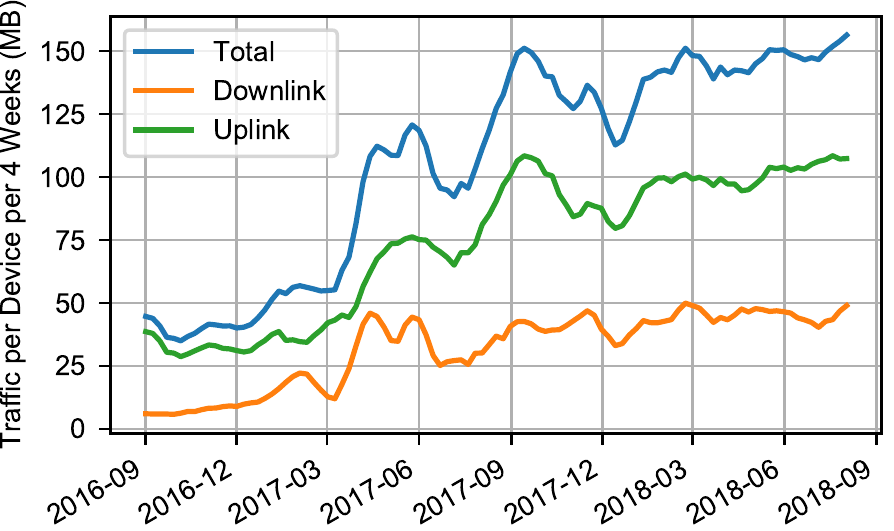}
\caption{Traffic per device per 4 week period}
\label{fig:data_per_imsi_two_years}
\end{figure}

First, we examine the traffic of cellular IoT devices over time to understand its evolution. Figure \ref{fig:data_per_imsi_two_years} shows the four-week moving average of traffic per device. We find that total IoT traffic per device increased three-fold, whereas downlink traffic increased six-fold. Most of the traffic growth occurred between September 2016 and 2017. Comparatively, \cite{shafiq2013} reported an total IoT traffic increase of 250\% during 2011. Furthermore, despite some fluctuations, the traffic does not demonstrate any seasonal patterns.

\begin{figure}
\centering
\includegraphics[width=\columnwidth]{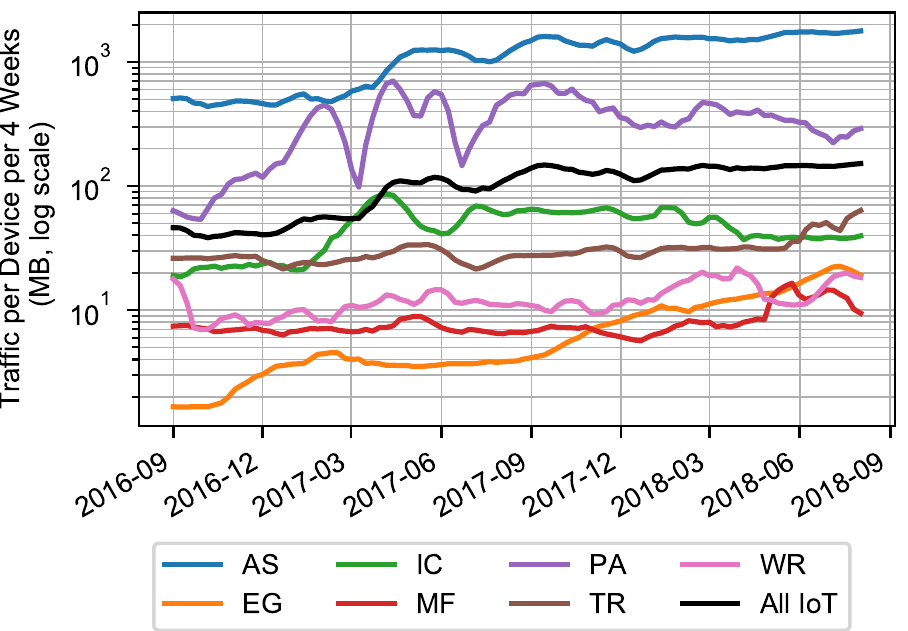}
\caption{Traffic per device per 4 week period for industries}
\label{fig:data_per_imsi_industry_two_years}
\end{figure}

Regarding different industries Figure \ref{fig:data_per_imsi_industry_two_years} shows the four-week moving average of traffic per device by industry. We observe significant differences in traffic volumes between industries, with devices in {\it Manufacturing} generating on average 10 MB per four weeks, while devices in {\it Administrative and support} (dominated by security companies) generating 2 GB per device, potentially due to security cameras generating video traffic. Furthermore, the volume of traffic increased in all industries. The most substantial increase occurred in {\it Electricity and gas} where traffic grew twelve-fold to 19 MB per device. We also analyzed the traffic evolution for the subset of companies that had active IoT devices in both the first and last months of the observation period and found similar trends. This indicates that the increase in traffic over time includes both companies that already use IoT and new companies adopting IoT.

\begin{figure}
\centering
\includegraphics[width=\columnwidth]{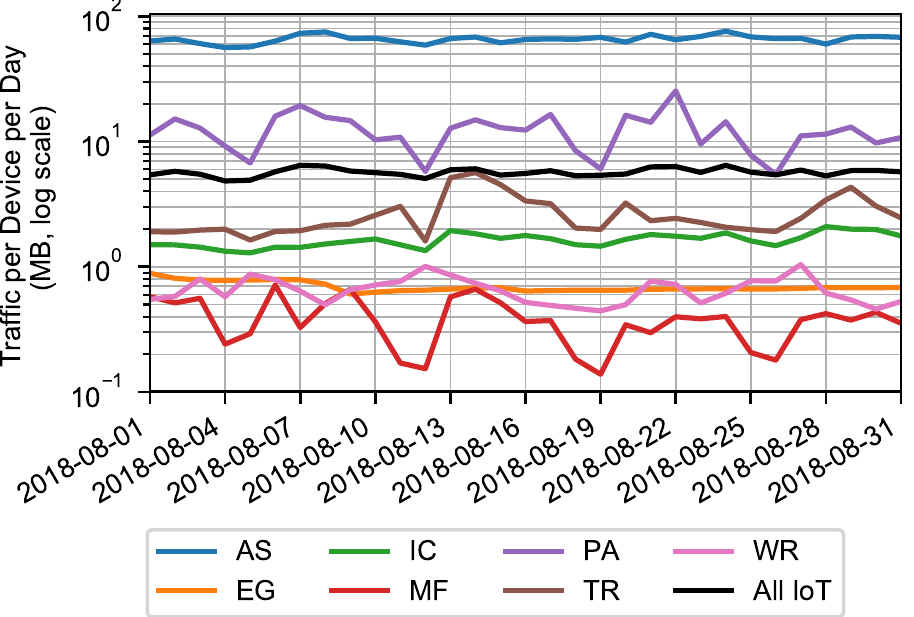}
\caption{Traffic per device per day for August 2018}
\label{fig:data_per_day}
\end{figure}

To explore differences in traffic depending on the day of the week, we study the daily traffic per device for August 2018. Figure \ref{fig:data_per_day} illustrates this traffic. We find that most industries do not show significant variation depending on the day of the week. However, in {\it Professional activities} and {\it Manufacturing} industries, we observe weekday-weekend patterns, with traffic halving during weekends.  

\begin{figure}
\centering
\includegraphics[width=\columnwidth]{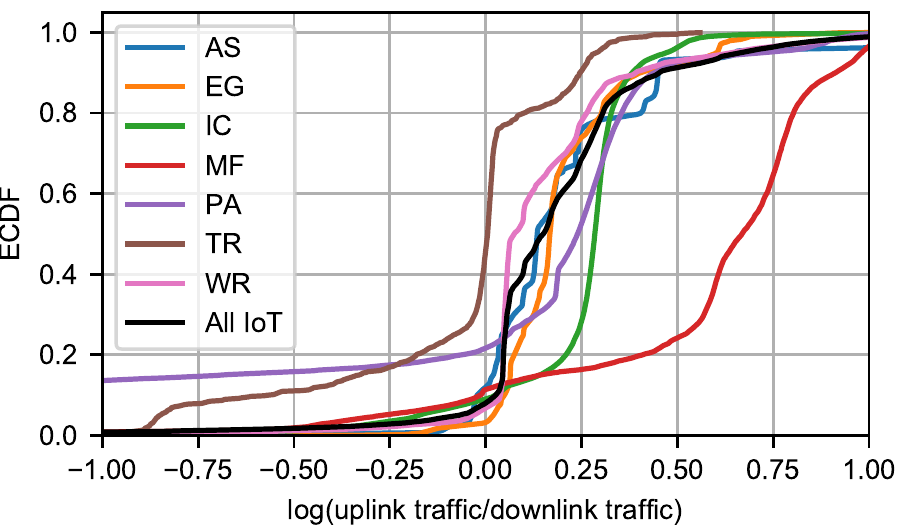}
\caption{ECDF of the log of uplink to downlink traffic ratio for August 2018}
\label{fig:Z_value}
\end{figure}
Furthermore, similar to \cite{shafiq2013}, we study uplink vs. downlink traffic volumes in IoT. Figure \ref{fig:Z_value} illustrates the empirical cumulative distribution (ECDF) of the log of uplink/downlink ratio for August 2018. Negative values indicate larger downlink traffic than uplink and positive values vice versa. As the figure shows, 92\% of IoT devices generated more uplink than downlink traffic, which is consistent with the finding of \cite{shafiq2013}, but exceeds the observation of \cite{Romirer-Maierhofer2015} by more than 30\%. However, in some industries, particularly {\it Transportation} and {\it Professional activities}, the share of devices with greater uplink than downlink traffic is lower, around 54\% and 78\% respectively. Further, in {\it Manufacturing}, the uplink traffic is much larger than downlink traffic (compared to other industries), with a median ratio of 4.66 compared to 1.41 for all IoT devices. Overall, the results illustrate both intra and inter-industry variation that helps illustrate the diversity of IoT.

\subsection{Mobility statistics}
\begin{figure}
\centering
\includegraphics[width=\columnwidth]{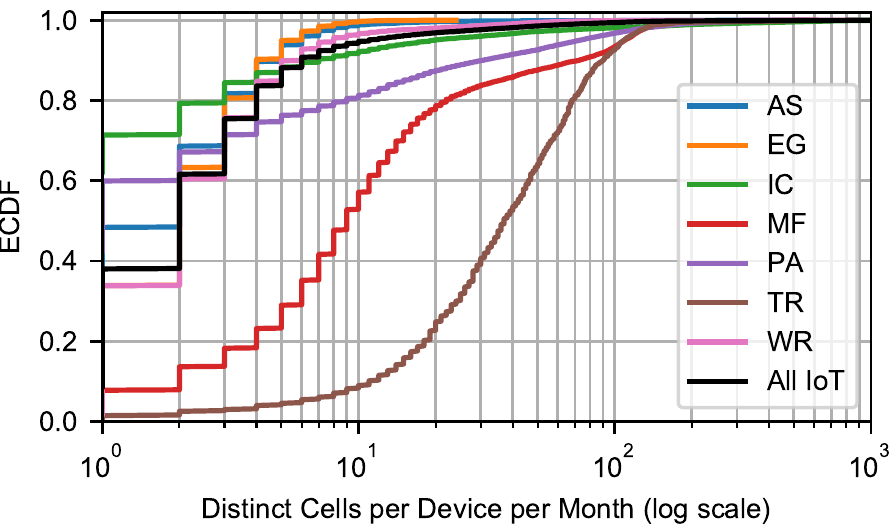}
\caption{ECDF of IoT device mobility for August 2018}
\label{fig:mobility_CDF}
\end{figure}

With regard to IoT device mobility, we infer mobility through the number of different cells visited\footnote{The definition of visit only includes cells where traffic was sent or received and thus is a lower bound on the number of cells attached to by the device.} by devices. Figure \ref{fig:mobility_CDF} presents the ECDF of the number of unique cells visited by devices in August 2018. As the figure shows, about one-third of IoT devices visited only a single cell, which is close to the 30\% rate observed by \cite{shafiq2013} for their one-week dataset. Furthermore, the share of stationary devices is likely higher since devices may visit different overlapping cells depending on varying signal strength. Around 95\% of all IoT devices and most devices in {\it Electricity and gas}, {\it Wholesale and retail trade}, and {\it Administrative and support} industries visited less than ten cells per month. Contrastingly and expectedly, devices in the {\it Transportation} industry are very mobile, with 90\% having visited more than ten cells. Some industries, for example {\it Manufacturing}, include a mix of mobile and stationary devices.

\subsection{Traffic and device distribution over cells}
\begin{figure}
\centering
\includegraphics[width=\columnwidth]{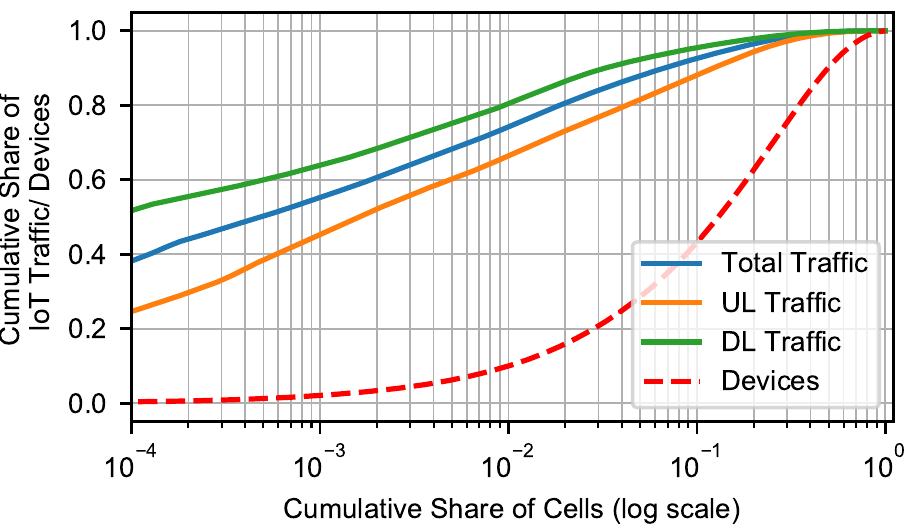}
\caption{Distribution of traffic and devices among cells for August 2018}
\label{fig:spatial_analysis}
\end{figure}

In addition to basic mobility, we can analyze the distribution of devices and traffic over all the cells IoT devices visited. Figure \ref{fig:spatial_analysis} illustrates the ECDF of IoT traffic and devices\footnote{We note that devices are counted once in each visited cell.} across IoT-visited cells in August 2018. The traffic is highly concentrated spatially, with 10\% of cells carrying about 93\% of total IoT traffic. Comparatively, \cite{paul2011} found 10\% of cells carrying about 55\% of total network traffic in a nationwide network in 2007. The high concentration of IoT devices and traffic is intuitive given the typical centralization of company campuses compared to normal consumers. This concentration is important for network planning because deployment of IoT-specific network features or optimizations would require changes to far fewer cells (and thus cost less) than for normal features. Finally, in terms of devices, we find that 10\% of cells account for about 44\% of devices while 50\% of cells about 90\% of devices, showing only moderate spatial concentration.

\subsection{Device base statistics}
\begin{figure}
\centering
\includegraphics[width=\columnwidth]{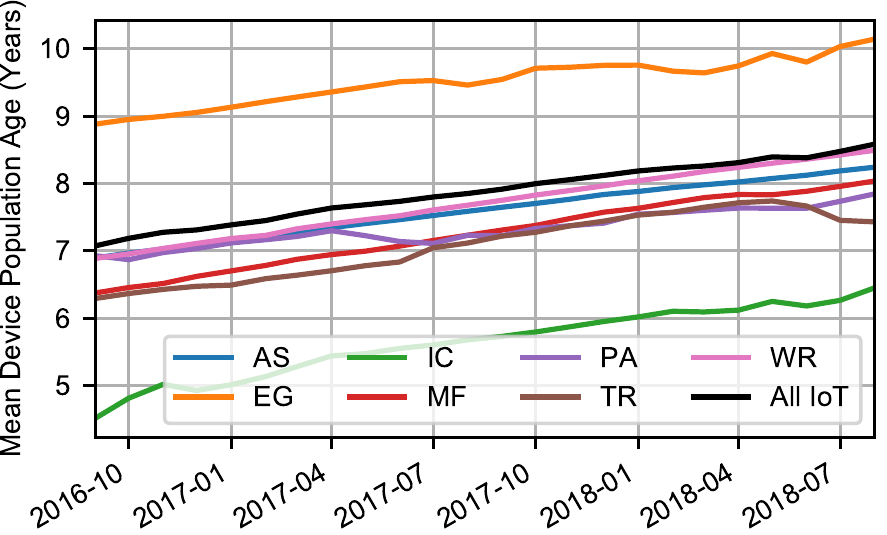}
\caption{Mean device population age (in years) based on the device model release year}
\label{fig:device_population_age}
\end{figure}

Through leveraging the additional device features dataset, we can analyze the feature and age evolution of the IoT device population. Figure \ref{fig:device_population_age} shows the evolution of the mean device population age by industry, with age defined as the time since the release year of the device model\footnote{We assume that device models are released on Jan. 1st. In other words, we overestimate the actual device population age, but this does not preclude tracking temporal dynamics and comparing industries.} of the device (and not the manufacturing year of the device). As Figure \ref{fig:device_population_age} shows, the mean IoT device population age as of August 2018 was above 8.5 years. The {\it Electricity and gas} industry has the oldest IoT device population, with a mean age of more than ten years as of August 2018. Overall, the increase in mean population age for all industries illustrates the slow pace of new device model deployment. 

\begin{figure}
\centering
\includegraphics[width=\columnwidth]{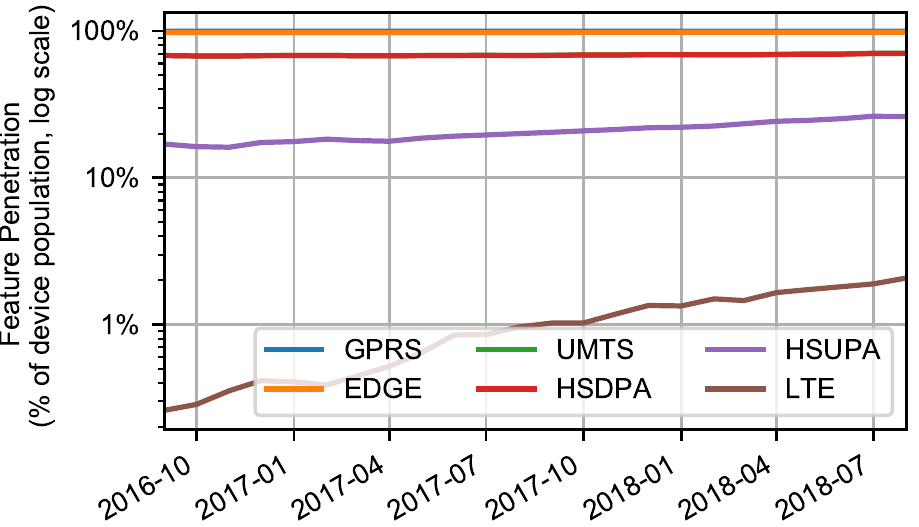}
\caption{Penetration of IoT device features}
\label{fig:feature_diffusion}
\end{figure}

In terms of connectivity features, Figure \ref{fig:feature_diffusion} presents the penetration of 3GPP connectivity technologies among the IoT device population. First, we observe the low penetration (and growth rate) of LTE of about 2\% as of August 2018. This observation is in line with the significant age of the IoT device population, and contrasts with LTE penetration of 41\% among non-IoT devices in Europe in 2017 \cite{GSMAssociation2018}. We further find that although the penetration of LTE is growing in all industries, only in the {\it Transportation} industry has penetration exceeded 10\%. Furthermore, we observe a significant difference in the penetration of HSDPA and HSUPA technologies of 70\% and 26\% respectively. This is surprising given the prevalence of uplink traffic in IoT which suggests a stronger need for fast uplink technologies rather than downlink. Finally, we find the share of 2G only (GPRS and EDGE) devices is still about 30\%. Therefore, discontinuing 2G service (for spectrum reuse purposes) would indeed affect a significant fraction of IoT devices thus posing a problem for network operators. 

\subsection{Temporal analysis}
For temporal insights, we perform several types of temporal analysis on different time scales. Specifically, our approach is to study the short timescale (hours, weeks) temporal patterns of three different months roughly evenly spaced over the two years: September 2016, August 2017, and August 2018. We always present the results from August 2018 and only present and note the results from the earlier months if substantially different.

\begin{figure}
\centering
\includegraphics[width=\columnwidth]{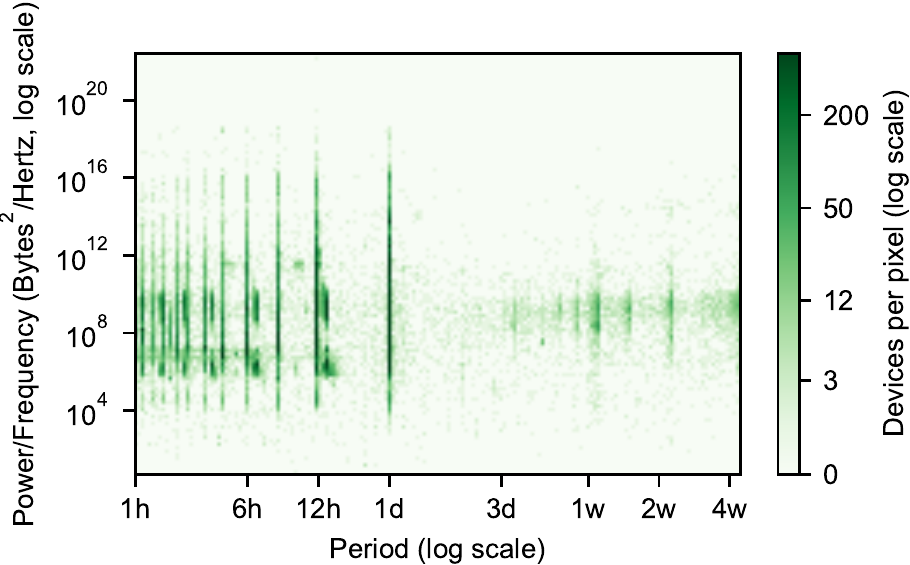}
\caption{Density of spectrum peaks vs periods of devices for total traffic for August 2018}
\label{fig:periodogram_data_in_peaks}
\end{figure}

First, we perform spectral analysis on a one-month traffic volume series of each device for uplink and downlink traffic. The spectral density of each series is estimated as the squared modulus of the Discrete Fourier Transform, in other words the periodogram. Then the peak power and corresponding period are extracted from each periodogram under the assumption that most IoT devices will have a dominant timer-driven peak. Figure \ref{fig:periodogram_data_in_peaks} illustrates the density of these (peak power, period) pairs for downlink traffic. The plot for uplink is almost identical. We find large fractions of devices have peaks at 24, 12, and 6 hour periods including devices with large and small peak traffic volumes (power). However, we also find other periods such as \texttildelow13 hours, though this case is specific to only two large companies with similar device models. The reason for the use of a 13 hour period in these companies is unknown. We also note that some devices have peaks at a 1 week period thus reinforcing the patterns from Figure \ref{fig:data_per_day}, however these devices tend to have small peak traffic volumes.

For a more detailed temporal analysis, we perform temporal clustering on the averaged (over the month) and normalized 24-hour total traffic volume series of each device. The normalization is performed for each device over the 24-hour series such that the value for any given hour is the fraction of that device's total daily traffic in that hour. This normalization is required due to the order of magnitude differences in traffic volumes between some devices. Each series is then transformed by a discrete wavelet transform (DWT) with a Daubechies-1 wavelet and a decomposition level of 3.

The DWT coefficients are then clustered via bisecting k-means with the number of clusters chosen by the silhouette score. We use bisecting k-means because of the $\mathcal{O}(n)$ run-time and ease of computational distribution. Comparatively, other approaches such as hierarchical clustering with ward linkage have a run-time of $\mathcal{O}(n^{2})$ \cite{karypis2000}. Though for robustness, we also cluster a random sample of 2000 devices via hierarchical clustering with ward linkage and with the number of clusters chosen by the Davies-Bouldin score. This is the same clustering setup as in \cite{shafiq2013}. We find the same number of clusters as the full device clustering and virtually the same cluster centroids.

\begin{figure}
\centering
\includegraphics[width=\columnwidth]{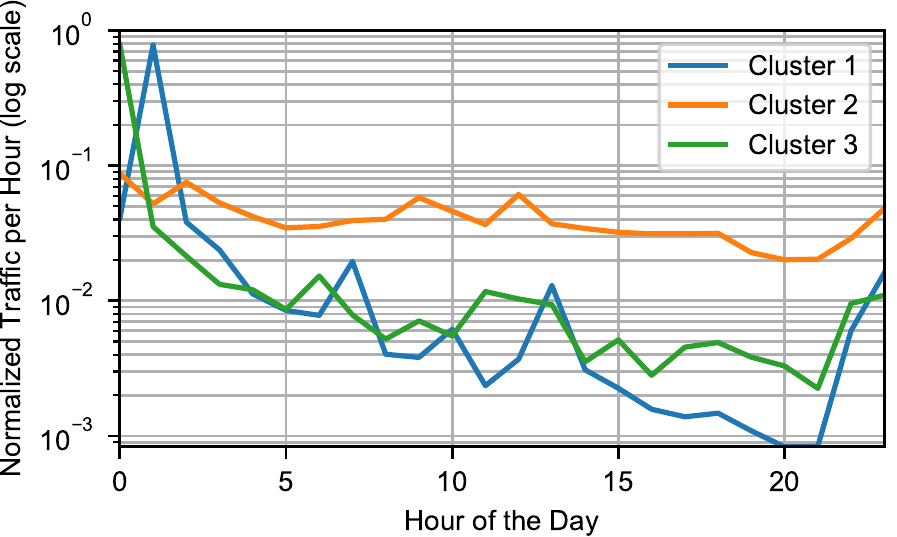}
\caption{Cluster centroids of the three temporal clusters from August 2018}
\label{fig:cluster_centroids}
\end{figure}

Regarding clustering results, we find that the optimal number of clusters is three. The clusters denoted 1, 2, and 3 encompass 25\%, 41\%, and 34\% of devices respectively. The cluster centroids (in terms of time series rather than DWT coefficients) of the three clusters are illustrated in Figure \ref{fig:cluster_centroids}. We find that clusters 1 and 3 have significant peaks at 0:00 and 2:00 respectively with over 80\% of their traffic within that peak hour, while cluster 2 shows much steadier and flatter traffic throughout the day.

To better understand these patterns we look at the composition of the clusters by company ID and industries. Interestingly, 88\% of cluster 1 devices belong to a single large company; thus this cluster is company-specific and not necessarily a general IoT temporal pattern. Though \cite{shafiq2013} also found an outlier cluster with a peak at 02:00 that they attributed mainly to fleet management applications. For clusters 2 and 3, no single company represents more than 31\% of devices and no single industry more than 50\% of devices. The main industries for cluster 2 are {\it Wholesale and retail trade} (40\%), {\it Electricity and gas} (22\%), and {\it Information and communication} (14\%), while the main industries for cluster 3 are {\it Wholesale and retail trade} (51\%), {\it Electricity and gas} (30\%), {\it Administrative and support services} (11\%). This overlap in industries highlights diversity in use cases even within narrow industries such as {\it Electricity and gas}.

\begin{figure}
\centering
\includegraphics[width=\columnwidth]{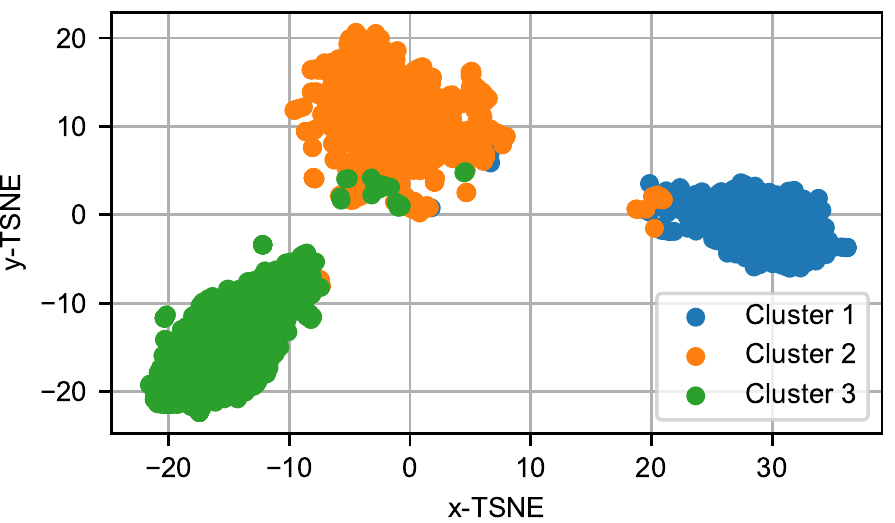}
\caption{T-SNE of sample of 4400 devices from the three temporal clusters from August 2018}
\label{fig:tsne_results}
\end{figure}

For an illustration of cluster separation, we plot the t-distributed stochastic neighbor embedding (t-SNE) of a random sample\footnote{t-SNE has a run-time complexity of $O(n^{2})$ and thus does not scale to large data.} of 4400 devices in Figure \ref{fig:tsne_results} with perplexity chosen as in \cite{cao2017}. The clusters appear to be well separated with only minimal overlap, especially the single-company dominated cluster 1, thus reinforcing the clustering results.

In terms of longer scale temporal phenomena, we did not find large differences in either temporal analysis method between the examined months. This suggests that IoT phenomena change slowly; such behavior reinforces the previously identified slow change in, for example, device feature penetration.

\section{Discussion}\label{sec:dicussion}
First, the study has several limitations that should be noted. The dataset is from only a single MNO in a single country and includes only industry IoT and not consumer IoT. Therefore, the study is not fully representative. However, we hope that similar studies from other countries and MNOs can help to build up a wide-ranging and practical understanding of IoT dynamics. Additionally, the time resolution of 1 hour means that the study might have missed more granular phenomenon on the minute and second timescales. However, we note that potentially more intrusive data collection methods would be required for those timescales, thus potentially hindering collection.

Overall, this work presented an analysis of cellular IoT traffic and mobility patterns over several different timescales for a major Finnish MNO. The analysis includes trends over a two-year span thus allowing a view of the evolution of IoT. Moreover, trends were broken down by industry, and the penetration of device features in the IoT device population was analyzed. Overall the analysis provided a diverse set of results of which we highlight a few.

For example, we found that IoT traffic per device tripled over the last two years, however the mean age of IoT device models in the device population also increased significantly to over eight years. Furthermore, the penetration of LTE-enabled IoT devices is very low (2\%) and growing very slowly. Also we found significant variation between devices of different industries with orders of magnitude differences in traffic volume and mobility. Finally, IoT devices can be clustered into only a few daily temporal patterns with many devices split between a relatively flat pattern and or a peaked pattern with a peak at midnight.

The results have implications for the design of cellular networks and IoT services. In particular, the results can help MNOs to understand differences in the IoT use across industries, which is necessary, for example, for defining virtual network slices to optimize service quality for specific use cases. Such network slicing is expected to become a key technology in future 5G networks. The trends in our study can also be used for forecasting of IoT traffic and device feature penetration. And finally, the results can help by grounding the IoT hype through empirical observations.

\section*{Acknowledgment}
This work has been supported by the Digital Disruption of Industry project funded by the Strategic Research Council (grant number: 292889) of Finland.
\bibliographystyle{IEEEtran}
\bibliography{ms}
\end{document}